\documentclass{emulateapj}

\newcommand{\myemail}{ts.kawaguti@nao.ac.jp}

\def\brfrac#1#2{\left(\dfrac{#1}{#2}\right)}
\def\dfrac#1#2{{\displaystyle\frac{\mathstrut #1}{#2}}}
\slugcomment{
Accepted for Publication in ApJL
}

\shorttitle{Wandering Black Hole and Accompanying Star Cluster 
in the M31 Halo}
\shortauthors{Kawaguchi et al. 2014}
\begin{document}
%%%%%%%%%%%%%%%%%%%%%%%%%%%%%%%%%%%%%%%%%%%%%%%%%%%%%%%%%%%%%%%%%%%%%%%%%%%%%%
%%%%%%%%%%%%%%%%%%%%%%%%%%%%%%%%%%%%%%%%%%%%%%%%%%%%%%%%%%%%%%%%%%%%%%%%%%%%%%

\title{
Relics of Galaxy Merging: Observational Predictions 
for a Wandering Massive Black Hole and Accompanying Star Cluster 
in the Halo of M31}

\author{Toshihiro Kawaguchi$^{1, 2}$, Yuriko Saito$^{3, 4}$, Yohei Miki$^{2}$, 
and Masao Mori$^{2}$}
\affil{$^1$
Astronomy Data Center, 
National Astronomical Observatory of Japan, 
Mitaka, Tokyo 181-8588, Japan; 
  \myemail}
\affil{$^2$Center for Computational Sciences, University of Tsukuba, 
  Tsukuba, Ibaraki 305-8577, Japan}
\affil{$^3$Department of Astronomical Science, 
The Graduate University for Advanced Studies (SOKENDAI), 
  Mitaka, Tokyo 181-8588, Japan}
\affil{$^4$
Subaru Telescope, National Astronomical Observatory of Japan, 
  650 North A'ohoku Place, Hilo, HI 96720, USA
}

%%%%%%%%%%%%%%%%%%%%%%%%%%%%%%%%%%%%%%%%%%%%%%%%%%%%%%%%%%%%%%%%%%%%%%%%%%%%%%
%%%%%%%%%%%%%%%%%%%%%%%%%%%%%%%%%%%%%%%%%%%%%%%%%%%%%%%%%%%%%%%%%%%%%%%%%%%%%%
\begin{abstract}
Galaxies and massive black holes (BHs) presumably 
grow via galactic merging 
events and subsequent BH coalescence. 
As a case study, we investigate 
the merging event between the Andromeda galaxy (M31) and a 
 satellite galaxy. 
We compute the expected observational appearance of 
the massive BH that was at the center of 
the satellite galaxy prior to the merger,  
and is currently wandering in
the M31 halo. 
We demonstrate that a radiatively inefficient accretion flow 
with a bolometric luminosity of a few tens of 
solar luminosities develops
when Hoyle-Lyttleton accretion onto 
the BH is assumed.
We compute the associated broadband spectrum 
and show 
that the radio band (observable with EVLA, ALMA and SKA) 
is the best 
frequency range to detect the emission. 
We also evaluate the mass and the luminosity of the stars bound by 
the wandering BH and find that such a star cluster 
is sufficiently luminous 
that it could correspond to one of the 
star clusters found by the PAndAS survey.
The discovery of a relic massive BH wandering in a galactic halo 
 will provide a direct means to investigate in detail the
coevolution of galaxies and BHs.
It also means a new population of BHs (off-center massive BHs), 
and offers 
targets for clean BH imaging that avoids strong interstellar scattering 
in the center of galaxies. 
\end{abstract}
%%%%%%%%%%%%%%%%%%%%%%%%%%%%%%%%%%%%%%%%%%%%%%%%%%%%%%%%%%%%%%%%%%%%%%%%%%%%%%
%%%%%%%%%%%%%%%%%%%%%%%%%%%%%%%%%%%%%%%%%%%%%%%%%%%%%%%%%%%%%%%%%%%%%%%%%%%%%%

%%%%%%%%%%%%%%%%%%%%%%%%%%%%%%%%%%%%%%%%%%%%%%%%%%%%%%%%%%%%%%%%%%%%%%%%%%%%%%
%%%%%%%%%%%%%%%%%%%%%%%%%%%%%%%%%%%%%%%%%%%%%%%%%%%%%%%%%%%%%%%%%%%%%%%%%%%%%%
%% Keywords should appear after the \end{abstract} command. The uncommented
%% example has been keyed in ApJ style. See the instructions to authors
%% for the journal to which you are submitting your paper to determine
%% what keyword punctuation is appropriate.

%% Authors who wish to have the most important objects in their paper
%% linked in the electronic edition to a data center may do so in the
%% subject header.  Objects should be in the appropriate "individual"
%% headers (e.g. quasars: individual, stars: individual, etc.) with the
%% additional provision that the total number of headers, including each
%% individual object, not exceed six.  The \objectname{} macro, and its
%% alias \object{}, is used to mark each object.  The macro takes the object
%% name as its primary argument.  This name will appear in the paper
%% and serve as the link's anchor in the electronic edition if the name
%% is recognized by the data centers.  The macro also takes an optional
%% argument in parentheses in cases where the data center identification
%% differs from what is to be printed in the paper.

% max=6
%%%%%%%%%%%%%%%%%%%%%%%%%%%%%%%%%%%%%%%%%%%%%%%%%%%%%%%%%%%%%%%%%%%%%%%%%%%%%%
%%%%%%%%%%%%%%%%%%%%%%%%%%%%%%%%%%%%%%%%%%%%%%%%%%%%%%%%%%%%%%%%%%%%%%%%%%%%%%

%%%%%%%%%%%%%%%%%%%%%%%%%%%%%%%%%%%%%%%%%%%%%%%%%%%%%%%%%%%%%%%%%%%%%%%%%%%%%%
%%%%%%%%%%%%%%%%%%%%%%%%%%%%%%%%%%%%%%%%%%%%%%%%%%%%%%%%%%%%%%%%%%%%%%%%%%%%%%
\section{Introduction}

In the hierarchical structure formation scenario in a 
cold dark matter universe, 
dark matter halos and their galaxies 
have grown and built up their mass
by colliding and merging with ambient sub-halos and satellite galaxies 
(e.g., Bullock \& Johnston 2005; Chiba et al. 2005). 
Furthermore, massive
galaxies harbour central, massive black holes (BHs), 
whose masses ($M_{\rm BH}$) are correlated with the 
mass of the spheroidal component 
of the host galaxy (Kormendy \& Richstone 1995). 
Currently, the BH--bulge relation is observed down to $M_{\rm BH}$ of 
$\sim 10^5 M_\odot$ 
(Barth et al. 2005; Xiao et al. 2011). 
One is therefore naturally led to believe that a massive 
BH also increases its mass 
after a galaxy collision by merging with 
the BH that was located at the center of 
a satellite galaxy, as presumed in semi-analytical 
 calculations for the galaxy and BH evolution 
(Kauffmann \& Haehnelt 2000).

There is, however, no clear
observational evidence that the merging events 
between galaxies really lead to the merging of massive BHs.
The central BH of a satellite galaxy likely
 wanders in the halo of its host galaxy after a galaxy merging event, 
before it finally sinks towards the center 
of the host galaxy due to the dynamical friction (Bellovary et al. 2010). 
Namely, 
the massive BH is expected to be located far away from 
the galactic center. 

There are some known 
implications for
the BH growth associated with 
merging events. 
For instance, NGC\,6240 (at $\sim$100\,Mpc) 
is an irregular starburst galaxy, indicative of 
a recent galaxy merging event,  
and harbours two massive BHs 
with a projected separation of 1\,kpc 
(Komossa et al. 2003).
Double or triple massive BHs in a colliding galaxy 
are now commonly found 
(e.g., Koss et al. 2011).
Moreover, 
ESO\,243--49, an edge-on disk galaxy at a similar distance, 
shows a bright X-ray 
point source (Hyper-luminous X-ray source)  
at a projected distance from the center 
of 3.5\,kpc (Farrell et al. 2009). 
A detailed X-ray spectral analysis 
revealed 
a $\sim 2 \times 10^4 M_\odot$ BH (Godet et al. 2012). 
Although this intermediate-mass BH candidate may 
have been 
the central BH of a satellite galaxy 
(Farrell et al. 2012), 
there is no clear evidence for a merging 
event in this galaxy.
In the case of some massive binary BH candidates  
(Sudou et al. 2003; Boroson \& Lauer 2009), 
their large distances 
make it difficult to detect 
merging signatures. 

As a case study, 
we focus on the nearby galaxy M\,31 (Andromeda galaxy). 
Its  proximity to the Galaxy, at a distance of 
780\,kpc (McConnachie et al. 2003; 
$1" \approx $4pc and 
$1\degr \approx $14\,kpc), 
allows detailed investigations into the 
faint stellar structures, showing that M\,31 
is evidently in the process of the galaxy collision 
(Ibata et al. 2001; Irwin et al. 2005; Fardal et al. 2007; 
Mori \& Rich 2008).
Thus, it is a unique laboratory to examine how  
galaxy evolution and BH growth take place.
Another advantage is the detectability: 
M\,31's proximity is essential in order to detect the faint
a low-luminous emission of a wandering BH. 

Some stellar substructures in the M31 halo 
were shown to be remnants of a minor merger about $1$ Gyr ago. 
In a previous paper 
(Miki et al. 2014, hereafter Paper I), 
we established the probable position of the associated wandering 
massive BH by performing $N$-body simulations. 
Finding this BH will provide us with clues to understand 
the coevolution of galaxies and massive BHs. 
If we really discover a massive BH wandering in a 
galactic halo, it indicates that 
many such wandering BHs, 
the remnants of satellite galaxies, in 
distant galaxies are missing due solely to 
sensitivity limits of current instruments.
Since all massive BHs so far have been found at the 
center of each galaxy, 
it opens up the possibility for the discovery of 
a new population of massive BHs.

In this Letter, 
we determine  
the most efficient waveband(s)
to detect a wandering massive BH of the satellite galaxy 
in the M\,31 halo. 
In the next section, 
the basic properties of the satellite galaxy 
and its central BH
are briefly described. 
Then, we present the expected 
emission from 
the wandering BH and 
from an assembly of stars bound by the BH.
We conclude with a summary and discussion of this study in \S~4.

\section{The Colliding Satellite galaxy and its  
Central Massive Black Hole}

We briefly summarize the basic properties of the 
satellite galaxy that is interacting with M\,31. 

As the satellite dwarf galaxy falls toward M\,31, it is 
elongated by the tidal force from M\,31, and eventually
destroyed 
following a series of pericentric passages.
Now, we see only the remnants of the satellite galaxy as 
a stellar stream (the so-called Andromeda stream) and two 
(western and northeastern) stellar shells 
(e.g., Ibata et al. 2001; Irwin et al. 2005). 
Detailed comparisons between observations and 
numerical simulations of the morphologies and the radial 
velocities of the 
stream and shells 
constrain the orbit, the mass 
and the concentration parameter of the satellite galaxy 
(Fardal et al. 2007; Mori \& Rich 2008; 
Paper I).

Via the metallicity--mass relation of dwarf galaxies, 
the lower limit on the stellar mass of the progenitor 
is known to be $5 \times 10^8 M_\odot$ (e.g., Mori \& Rich 2008).
An upper limit on the dynamical mass 
of the progenitor ($\le 5 \times 10^9 M_\odot$) comes from 
the disk thickness of M\,31 (Mori \& Rich 2008). 

The BH-bulge mass 
relation suggests that the progenitor dwarf galaxy 
has a massive BH with $M_\mathrm{BH}$ of about $1/500$ 
of the mass of its host spheroidal component 
(Magorrian et al. 1998; Marconi \& Hunt 2003).
This raises the question of whether the 
colliding satellite galaxy did indeed have a central BH. 
Font et al. (2006) estimated the luminosity of the colliding galaxy to be 
$M_B \approx -17$\,mag ($L_B \approx 10^9 L_\odot$).
Adopting the $B$--$I$ colors of elliptical and spiral galaxies (Fukugita et al. 1995), 
this corresponds to $M_I$ of $-19.3$ and $-18.7$\,mag, respectively.
Galaxies around the current low-mass end in the 
BH-bulge mass relation 
are of similar or even fainter luminosities: For example,
POX\,52, a galaxy harboring a $\sim 10^5 M_{\odot}$ BH, 
has a dwarf elliptical morphology with $M_I = -18.4$\,mag 
(Thornton et al. 2008). 
Host galaxies of $10^{5-6} M_{\odot}$ BHs 
have $-23 < M_I < -18.8$\,mag (Jiang et al. 2011).
Thus, we suspect that the colliding satellite galaxy 
had a massive BH at its center with $M_\mathrm{BH}$ of (at most) 
about $10^7 M_{\odot}$
(i.e., 1/500 of $5 \times 10^9 M_{\odot}$).
We describe the $M_\mathrm{BH}$-dependency of the results later. 

If there was a central massive BH in the satellite galaxy, 
then where is the BH now? 
Based on 
the $N$-body simulations in Paper I, which involved an intensive 
survey of parameters relevant to the collision followed up by 
high-resolution calculations, 
we identified five orbit models that reproduce the stream 
and the two shells well.
The current position of the BH is then constrained to a small region 
($\sim \, 0.\degr6 \times 0.\degr7$) in the eastern outskirts of the M31 disk.
The expected proper motion 
is 33--43\,$\mu$-arcsec\,yr$^{-1}$.
The distance of the BH from the centre of M\,31 is 18--49\,kpc, 
with a projected distance on the sky plane of 0.4--0.8\,deg.

\section{Searching for the wandering BH}

We now focus on the emission from the wandering massive BH
and trailing stars. 

%%%%%%%%%%%%%%%%%%%%%%%%%%%%%%%%%%%%%%%%
\begin{figure}
  \begin{center}
    \epsscale{1.15}
    \plotone{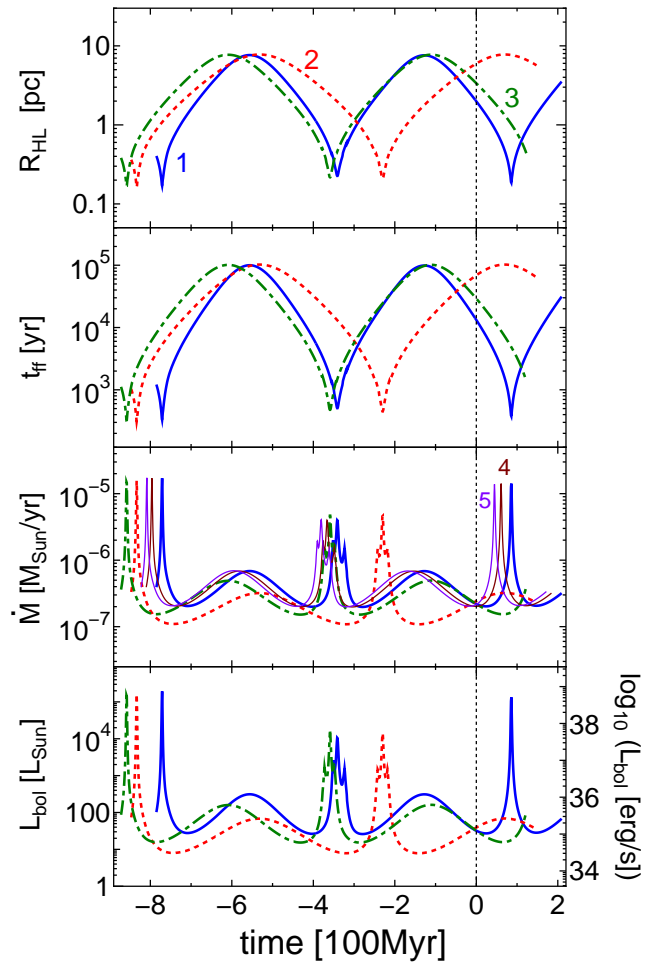}
    \caption{
Time variation of the 
Holye-Lyttleton radius $R_{\rm HL}$, 
the free-fall timescale at $R_{\rm HL}$, 
the accretion rate $\dot{M}$ and the bolometric luminosity 
for $M_{\rm BH} = 10^7 M_\odot$.
Blue solid, red dashed and green dot-dashed lines 
indicate the successful orbit models with ID\,1-3 in Paper I.
For $\dot{M}$, we also plot the results for ID\,4 and 5 by 
thin solid lines in brown and purple, respectively.
For different BH masses, 
$R_{\rm HL} \propto M_{\rm BH}$, 
$t_{\rm ff} \propto M_{\rm BH}$, 
$\dot{M} \propto M_{\rm BH}^2$
and  $L_{\rm bol} \propto M_{\rm BH}^3$.
The vertical dotted line indicates the present time.
        }
    \label{fig:mdot}
  \end{center}
\end{figure}
% Lsun = 3.85 x 10^33 erg/s = 10^33.59 erg/s
%%%%%%%%%%%%%%%%%%%%%%%%%%%%%%%%%%%%%%%%

\subsection{Hoyle-Lyttleton-Bondi accretion onto the BH}

First, we estimate the accretion rate onto the wandering massive 
BH.
Such a BH, moving with velocity of $v$, 
will capture 
the surrounding interstellar medium (ISM) of 
density $\rho$ and sound speed $c_s$ via   
Hoyle-Lyttleton-Bondi accretion at a rate $\dot{M}$
(Hoyle \& Lyttleton 1939; Bondi \& Hoyle 1944):
\begin{equation}
\dot{M} = 4 \pi G^2
\frac{\rho M_{\rm BH}^2}{(v^2 + c_s^2)^{1.5}}.
 \label{eq:mdot}
\end{equation}
We assume that the ISM has the same density profile as the mass 
profile of M\,31, 
namely, a Hernquist bulge, an exponential disk and 
an NFW halo (e.g., Fardal et al. 2007).
A constant gas fraction (with respect to the total mass 
of dark matter$+$star$+$gas) 
of 0.1 is adopted.
For the sound speed, 
we assume an ISM temperature 
of $10^4$\,K for the disk and bulge gas, and of $10^6$\,K for the 
halo gas.

The upper three panels of Figure~\ref{fig:mdot} show the 
Hoyle-Lyttleton radius $R_{\rm HL}$ [$\equiv 2 G M_{\rm BH} / (v^2 + c_s^2)$], 
the free-fall 
timescale at that radius 
($R_{\rm HL}^{1.5} / \sqrt{GM_{\rm BH}}$), 
and $\dot{M}$.
Among the five orbit models that reproduce the stellar structures of M\,31, 
the representative three cases (with ID\,1, 2 and 3) are drawn 
in all the panels for a BH mass of $10^7 M_\odot$.
Since the remaining two cases (ID\,4 and 5) show 
time variations similar to ID\,1, their results are 
drawn only in the third panel by thin solid lines.
To take into account different BH masses, 
each curve can be scaled as, 
$R_{\rm HL} \propto M_{\rm BH}$, 
$t_{\rm ff} \propto M_{\rm BH}$ and 
$\dot{M} \propto M_{\rm BH}^2$.
The time coordinate 
is shown with respect to the present time, 
so that negative values mean the past, and vice versa.

Around the pericenters 
(e.g., at $-770,\ -340$ and $+86$\,Myr for ID\,1), 
$R_{\rm HL}$ 
decreases due to the large BH velocity, 
whereas it has maxima around the apocenters 
(at $-560$ and $-130$\,Myr for ID\,1).
There are two types of peaks in the time variation of $\dot{M}$.
At the pericenters, the high values of $\dot{M}$ are due to 
the high ISM density 
near the M31 center, lasting only a
relatively short time because of the large BH velocity.
On the other hand, broad peaks in $\dot{M}$ around the 
apocenters are caused by small BH velocities.

To be precise, 
the estimated $\dot{M}$ is the mass capturing 
rate, which does not necessarily equal the mass accretion rate onto
the BH.
We now evaluate the 
outer radius  
of the accretion 
disk/flow, which is determined by the angular momentum of 
the captured ISM.
At the apocenter, 
$R_{\rm HL}$ is largest at about 10\,pc, i.e., 
10$^7$ Schwarzshild radii ($R_{\rm Sch}$) for a $10^7 M_{\odot}$ BH.
The ISM kinematics is mostly isotropic, but has a tiny anisotropy which forms 
the accretion disk.
The ISM captured at that distance from the BH will have 
a specific angular momentum of about 
5\,km\,s$^{-1}$\,pc (Phillips 1999). 
This corresponds to the Keplerian angular momentum at 
$\sim 500 R_{\rm Sch}$ 
for a $10^7 M_{\odot}$ BH.
Due to the redistribution of the angular momentum within the flow, 
the disk is a few times larger than 
the angular momentum barrier  
(e.g., Machida et al. 2004). 
Thus, we assume hereafter the size of the accretion flow 
to be $1000 R_{\rm Sch}$.
Therefore, the accretion timescale within the flow 
($\sim$ the free-fall timescale from the outermost radius 
divided by the viscosity parameter; Narayan \& Yi 1995)
is negligible compared with that 
from $R_{\rm HL}$ to the outermost 
radius of the flow.
It follows that the time required for the captured gas to reach %the vicinity of 
the BH is largely governed by the free-fall timescale at $R_{\rm HL}$, 
which is $\sim 10^5$\,yr at most (Figure 1), 
and is small compared to the orbital timescale 
of the wandering BH (some $10^8$\,yrs). 

The expected 
$\dot{M}$ is much smaller than the Eddington rate 
($\sim 10 L_{\rm Edd}/c^2 \approx 0.2 M_\odot$\,yr$^{-1}$ 
for a $10^7 M_\odot$ BH) by several orders of magnitude.
We, therefore, adopt an accretion model for such 
low rates, the 
advection-dominated accretion flow (ADAF) model. 
By imposing a radiative efficiency (the conversion efficiency 
from the rest mass energy of the infalling gas to the 
emergent radiation) of $\dot{M}/(L_{\rm Edd}/c^2)$ (Narayan \& Yi 1995), 
we calculate the bolometric 
luminosity as a function of time 
(bottom panel of Figure \ref{fig:mdot}).
The bolometric luminosity is a few tens of 
$L_\odot$ at the present time, and at most  
$\sim 10^{(4-5)} \, L_\odot$ around the pericenters.

\subsection{Radiation from the accretion flow}

%Fig: nu Lnu v.s. nu (ADAF, star cluster)
%%%%%%%%%%%%%%%%%%%%%%%%%%%%%%%%%%%%%%%%
\begin{figure*}
  \begin{center}
    \epsscale{1.0}
    \plotone{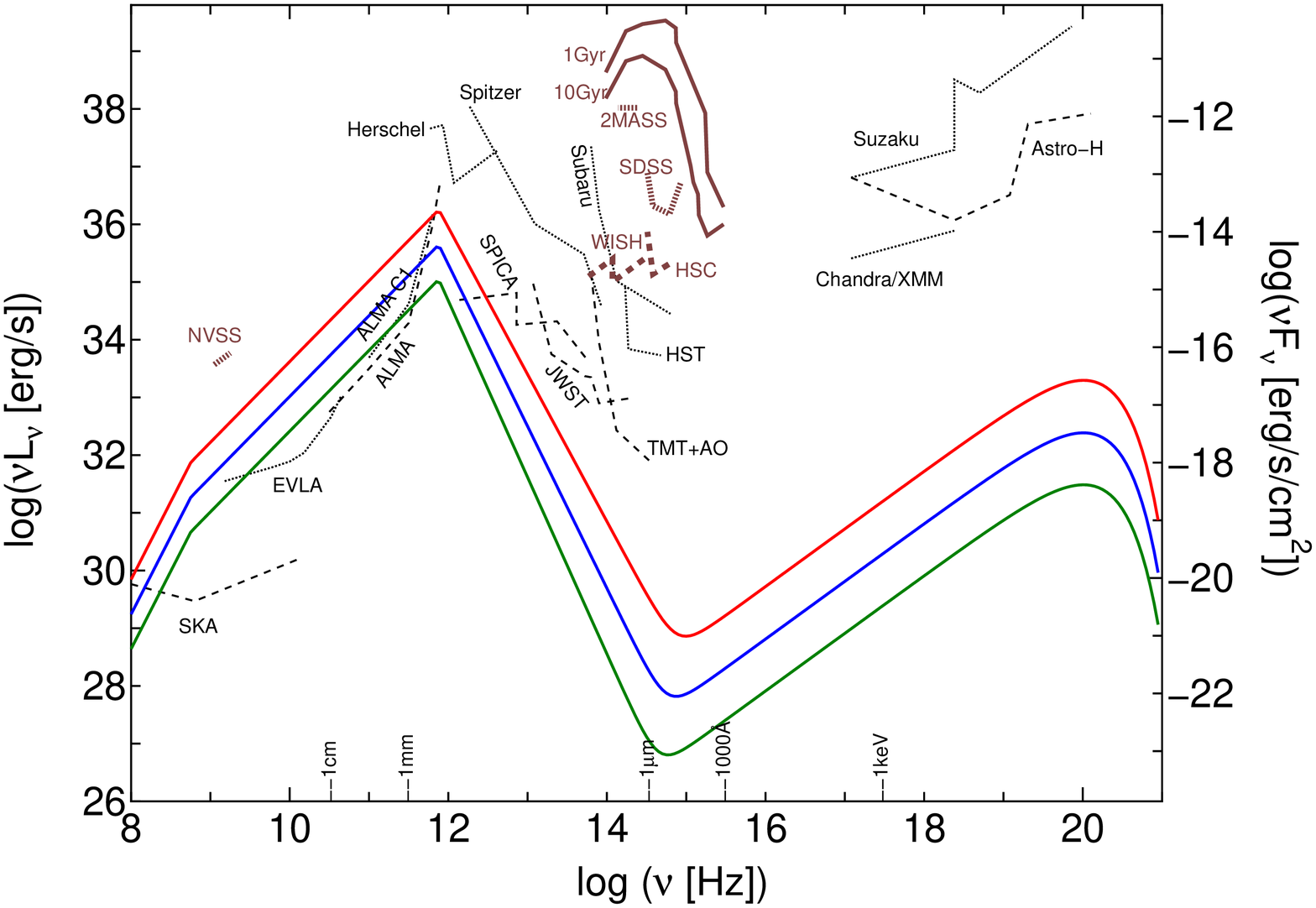}
    \caption{
      Expected spectral energy distribution of the 
wandering BH (solid curves)  
for three different 
BH masses [$5 \times 10^6$ (green), $10^7$ (blue) 
and $2 \times 10^7 M_\odot$ (red), 
 with $L_{\rm bol}$ of 4, 32 and 260 $L_\odot$, 
from the bottom to the top, respectively].
Right axis corresponds to the flux at the distance of M31. 
Black dotted and dashed 
lines indicate the  
detection limits (10$\sigma$ in $10^4$\,sec integration) 
of the existing and planned facilities, respectively.
The emission from the accretion flow 
is detectable only in 
the radio band (frequency less than 
$10^{12}$Hz).
Brown, thick solid curves are spectra of a star cluster that are 
1\,Gyr and 10\,Gyr old.
Brown, thick dotted and dashed lines indicate the 
10$\sigma$ sensitivities of wide-field surveys. 
    }
    \label{fig:seds}
  \end{center}
\end{figure*}
%%%%%%%%%%%%%%%%%%%%%%%%%%%%%%%%%%%%%%%%

Based on the current value of $\dot{M}$ shown in Figure~\ref{fig:mdot}, 
we present the expected broadband spectral energy distribution 
of the accretion flow in Figure~\ref{fig:seds}, 
using a simplified ADAF model (Mahadevan 1997).
Since the five successful orbit models exhibit almost 
the same $\dot{M}$, we only show the case for ID\,1.
Below $\sim 10^{12}$\,Hz, self-absorbed synchrotron emission emerges. 
The outermost radius ($10^3\,R_{\rm Sch}$)  
contributes at $\sim 10^{8.7}$\,Hz,  
while the innermost region at $3\,R_{\rm Sch}$ has a peak at 
$\sim 10^{12}$\,Hz.
Above that frequency, inverse Compton scattering of the radio 
seed photons arises up to $1\,\mu m$.
Bremsstrahlung emission appears at higher frequencies.

The detection limits (10$\sigma$ in $10^4$\,sec integration; 
based on, e.g., Perley et al. 2011 for EVLA and 
the ALMA sensitivity 
Calculator\footnote{http://almascience.nao.ac.jp/documents-and-tools}) 
of various instruments for a point source
are shown by solid and dotted lines in black.
The size of the accretion flow is about $10^{-3}$\,pc 
($10^3\,R_{\rm Sch}$ 
for a $10^7 M_{\odot}$ BH), corresponding to 
0.3\,milli-arcsec.
Thus, the flow is practically a point source for those facilities.

Because the emission from the BH 
is detectable only at 
frequencies less than $10^{12}$Hz,
 radio observations with EVLA, ALMA and SKA are crucial to 
find the wandering BH (see also Strader et al. 2012).
Since the luminosity from the self-absorbed synchrotron 
emission scales as $\propto M_{\rm BH}^{2/5} \, \dot{M}^{4/5} 
\propto M_{\rm BH}^2$ (Mahadevan 1997), 
the minimum detectable BH mass 
$\approx 3 \times 10^5 M_\odot$ for SKA.
If the accretion rate is much smaller than that in equation (1) 
by, e.g., 2.5\,dex, 
as is suggested by Perna et al.\ (2003) and Bower et al.\ (2003), 
the minimum detectable BH mass increases to about $3 \times 10^6 M_\odot$.
 Winds from stars surrounding 
the BH (\S\,3.3) could increase the accretion rate 
and thereby the luminosity of the flow.
Radio flares, as seen in Sgr\,A$^*$ (Miyazaki et al. 2004), 
will make the detection even easier.
The radio source 37W207 (Walterbos et al. 1985), located 
within our predicted region, may actually be the wandering BH.

\subsection{Star cluster}

The wandering BH likely trails stars that resided in the central
region of the satellite galaxy [see O'Leary \& Loeb (2009) 
and Merritt et al.\ (2009) for star clusters surrounding recoiled BHs].
The Hill radius $R_{\rm Hill}$ at the pericentric passage 
(with the minimum distance $R_0$ between the satellite and the M31 centers
of 1.1\,kpc) determines how many stars are still bound to the BH.
The M\,31 bulge (with a mass $M_{\rm bulge}$ of $3 \times 10^{10} M_\odot$ 
and a scale length 
$a$ of 0.6\,kpc) will straggle stars outside $R_{\rm Hill}$
from the BH via the tidal force, where 
\begin{eqnarray}
R_{\rm Hill} &=& 
\brfrac{M_{\rm BH}}{3 M_{\rm bulge}}^\frac{1}{3} R_0 \left( 1+\frac{a}{R_0}\right)^\frac{2}{3}\\
&\approx & 70\,{\rm pc}\brfrac{M_{\rm BH}}{10^7 M_\odot}^\frac{1}{3}.
 \label{eq:rhill}
\end{eqnarray}
The mass of stars in the satellite galaxy within this radius 
($\approx$\,18\,arcsec), if distributed in a King profile (see Paper I, 
for more details), amounts to 
about 10\% of the BH mass of the satellite galaxy.
If the presence of the central BH enhances the nuclear 
stellar density 
compared to that predicted by the King profile, 
the expected mass and luminosity of 
the star cluster will be higher, leading to an even easier detection.

Next, we utilize the 
stellar synthesis model PEGASE (Fioc \& Rocca-Volmerange 1997) to calculate 
the spectra of an ensemble of stars 
for various ages after an instantaneous starburst (their Figure\,1). 
In Figure\,\ref{fig:seds}, we draw such spectra for a star cluster of 
$10^6 M_\odot$ in stellar mass and ages of 
1\,Gyr 
and 10\,Gyr. 
The expected radiation is well above the 10$\sigma$ 
sensitivities of 
Near-infrared (IR) and optical, wide-field surveys, 
such as SDSS, HyperSuprime-Cam and WISH.
In fact, the PAndAS project (Mackey et al. 2010)
showed that there are several star clusters within the 
region Miki et al. (Paper I) proposed for the possible location of the wandering BH.
We propose 
that one of those clusters is the trailing star cluster that 
has survived against tidal destruction by M\,31.
Such remnant star clusters are likely more metal-enriched than normal
globular clusters 
[as is expected for stars surrounding recoiled BHs (Merritt et al.\ 2009)], 
since they consist of stars born
in the deep gravitational potential of the colliding
galaxies. 
If one obtains an indication of metal richness via
 red color and/or spectral features/indices, then such
 star clusters are good candidates for remnant clusters
 with massive BHs inside.

As another example, 
at the position of HLX-1  
of ESO\,243-49, there is a Near-IR, optical and ultraviolet 
counterpart with
 a brightness typical of a massive globular cluster  
(Soria et al. 2010; Farrell et al. 2012).
This may be another case of a BH wandering 
outside the galactic center 
and trailing a star cluster 
(see also Annibali et al. 2012; Bekki \& Chiba 2004).

\section{Summary and discussions}

We have investigated the observational signatures of 
a massive BH and associated star cluster relic 
from a late stage of galactic 
merging in the nearby galaxy M\,31. 
M\,31 likely experienced a merging event with a satellite galaxy about 
1\,Gyr ago, whose remnants are now observed as stellar sub-structures.

\begin{figure}
%  \begin{center}
    \epsscale{1.15}
    \plotone{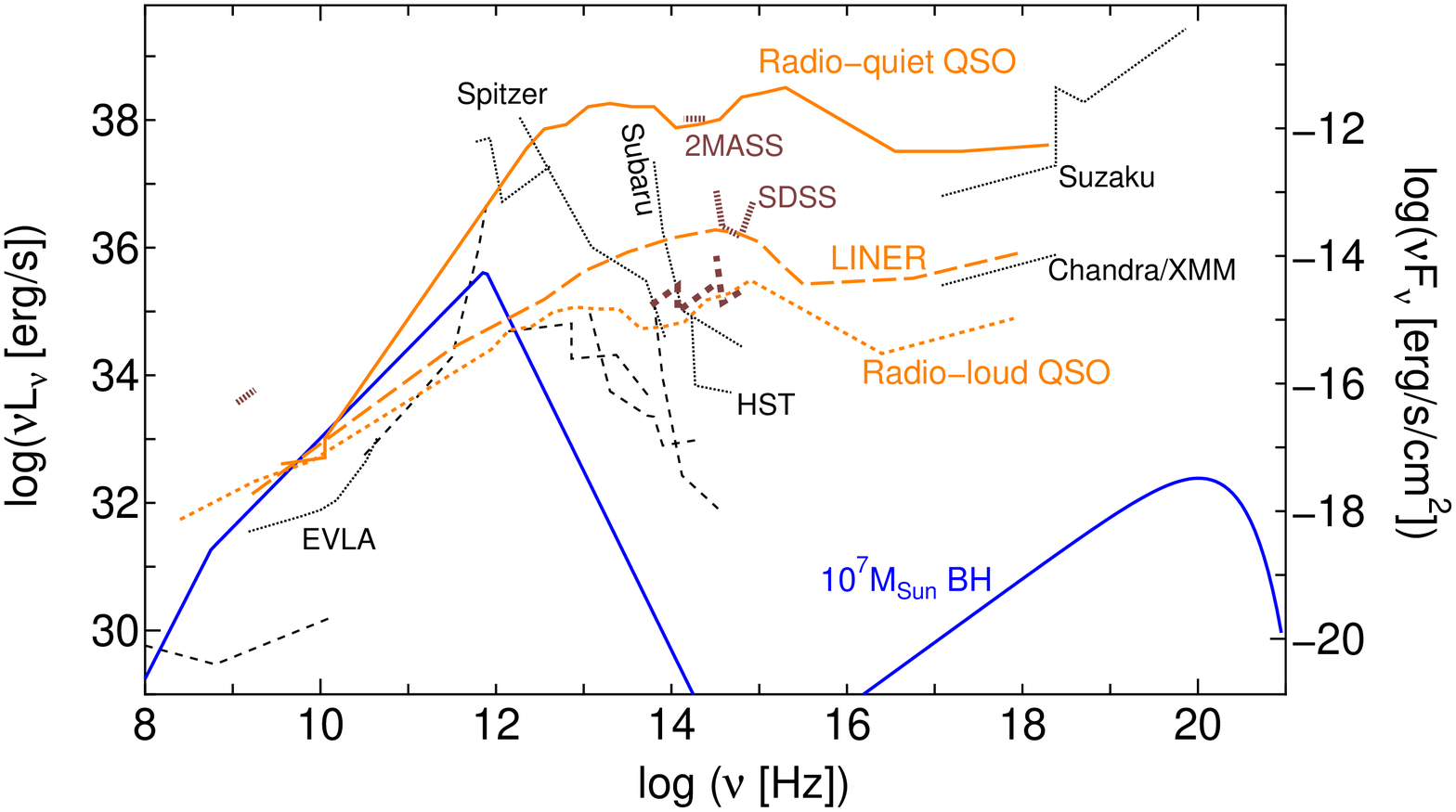}
    \caption{
Comparison of the spectral energy distributions
between our target (blue solid line for a 
$10^7 \, M_\odot$ BH), 
a radio-loud QSO at $z\sim 3$, 
a radio-quiet QSO at $z\sim 0.6$ and 
a LINER at $z\sim 0.02$ (orange dotted, solid and 
dashed lines, respectively; Nemmen et al. 2010).
The distances are chosen so that they have 
similar fluxes at 5\,GHz.
Detection limits are the same as those in Fig.~\ref{fig:seds}.
        }
    \label{fig:sed_other}
%  \end{center}
\end{figure}

Based on the expected position and velocity of the wandering BH 
that was originally at the center of the satellite galaxy, 
we have estimated the 
current gas accretion rate onto the BH via the Hoyle-Lyttleton-Bondi formula.
We then calculated the broadband spectrum using the estimated accretion 
rate adopting an ADAF model.
We find that  
the emission from the 
accretion flow is well above the sensitivities of the radio facilities 
currently operating (EVLA and ALMA Cycle 1).
Once the full ALMA and the SKA become operational, 
we will be able to probe the multi-waveband properties. 
As the spectral energy distributions of 
radio-loud,
radio-quiet active galactic nuclei, 
and LINERs are very different from that of our
target (Figure~\ref{fig:sed_other}), 
we can discriminate such background sources
from the accretion flow in other wavebands, 
including the IR, optical and X-ray.

%%%%%%%%%%%%%%%%%%%%%%%%%%%%%%%%%%%%%%%%
%%%%%%%%%%%%%%%%%%%%%%%%%%%%%%%%%%%%%%%%

Another potential approach to search for the wandering massive BH 
is to hunt for the stars that were in the central region of 
the satellite galaxy and  are, thus, likely to be more metal-enriched than 
normal globular clusters. 
The stars trailed by the BH 
would be easily detected by Near-IR and optical wide-field surveys.
We propose 
that one of the star clusters found by the PAndAS project 
in the area we constrained for the position 
of the BH, is the star cluster accompanying the wandering BH.

If we indeed detect such a 
wandering massive BH associated with the galaxy merging event, 
it will be an important contribution to 
exploring how the coevolution of BHs and galaxies proceeds.
A discovery will imply a new population of BHs; off-center 
massive BHs in galactic halos.
Recoiling BHs, kicked by the gravitational wave 
emission during a BH-BH merging event, 
will have a similar broadband spectral distribution 
and, thus, a similar radio detectability.
Moreover, such objects will enable us to image the accretion flow  
around a BH (Luminet 1979; Fukue \& Yokoyama 1988), 
which is practically very difficult for  
central BHs due to interstellar scattering in the centers of galaxies
%(Lo et al. 1999; 
(Doeleman et al. 2001).

%%%%%%%%%%%%%%%%%%%%%%%%%%%%%%%%%%%%%%%%%%%%%%%%%%%%%%%%%%%%%%%%%%%%%%%%%%%%%%
%%%%%%%%%%%%%%%%%%%%%%%%%%%%%%%%%%%%%%%%%%%%%%%%%%%%%%%%%%%%%%%%%%%%%%%%%%%%%%
\acknowledgments

We are grateful to 
the anonymous referee for detailed comments, 
and N.\ Nakai,   
M.\ Saito, I.\ Iwata and K.\ Ebisawa 
for useful discussions, 
and to Alex Wagner  
for careful reading of the manuscript.
This work was supported in part by the FIRST project based on 
Grants-in-Aid for Specially Promoted Research by MEXT (16002003) 
and Grant-in-Aid for Scientific Research (S) by JSPS (20224002), 
(A) by JSPS (21244013), and (C) by JSPS (25400222). 
\vspace{1ex}
%%%%%%%%%%%%%%%%%%%%%%%%%%%%%%%%%%%%%%%%%%%%%%%%%%%%%%%%%%%%%%%%%%%%%%%%%%%%%%
%%%%%%%%%%%%%%%%%%%%%%%%%%%%%%%%%%%%%%%%%%%%%%%%%%%%%%%%%%%%%%%%%%%%%%%%%%%%%%

%%%%%%%%%%%%%%%%%%%%%%%%%%%%%%%%%%%%%%%%%%%%%%%%%%%%%%%%%%%%%%%%%%%%%%%%%%%%%%
%%%%%%%%%%%%%%%%%%%%%%%%%%%%%%%%%%%%%%%%%%%%%%%%%%%%%%%%%%%%%%%%%%%%%%%%%%%%%%

%
% Figures
%

\end{document}